# Demo or Practice: Critical Analysis of the Language/Action Perspective

Mark Dumay

*Department of Information Systems & Software Engineering, Delft University of Technology, Delft, The Netherlands.*

**Abstract**

Despite offering several promising concepts, the Language/Action Perspective (LAP) is still not in the mainstream of Information Systems Development (ISD). Since at present there is only a limited understanding of LAP theory and practice, it remains unclear whether the lack of LAP's impact is due to shortcomings in LAP theory itself. One classic problem within ISD is the dichotomy between social perspectives and technical perspectives. LAP claims it offers a solution to this problem. This paper investigates this claim as a means to review LAP theory. To provide a structure to a critical analysis of DEMO – an example methodology that belongs to the LAP research community – this paper utilizes a paradigmatic framework. This framework is augmented by the opinion of several DEMO practitioners by means of an expert discussion. With use of a comparative evaluation of LAP theory and DEMO theory, the implication of DEMO's reflection upon LAP is determined. The paper concludes by outlining an agenda for further research if LAP is to improve its footprint in the field.

## 1  Introduction

This paper focuses on a classical problem in Information Systems Development, being the apparent dichotomy between social perspectives and technical perspectives. The Language/Action Perspective is an approach that is based upon analysis of communication as a basis for the design of Information Systems. This paper investigates if this approach can unify the dichotomy. This section elaborates on the setup of research that supports this investigation.

### 1.1  Research Background

Since its first application in the early 1950s and 1960s, Information Technology has had an increasing impact on organizations (Daft 1998). The notion of virtual organization – a form of organization that is no longer bound to physical presence by virtue of communications technology – is just one example of how such technology can change the foundation of modern organization (Jaffee 2001: 201-6). To stress that the application of Information Technology at an organizational scope has a profound impact on its context, such application is commonly known as an Information System. A classic dichotomy in the discipline of Information Systems Development (ISD) is to view such systems from either a technical perspective or from a social perspective (Goldkuhl and Lyytinen 1982). Nevertheless, Information Systems Development has to deal with both social and technical aspects (Hirschheim, Klein et al. 1995). The inherent contradiction of perspectives is an important cause of the failure of many Information Systems (cf. Riesewijk and Warmerdam 1988; Nuseibeh and Easterbrook 2000).

Most practitioners within the field of ISD are guided by the philosophical assumptions of functionalism (Goles and Hirschheim 2000). ISD research has a similar bias, although most notably the work of Goles, Hirschheim, Iivari, Klein and Lyytinen (Hirschheim, Klein et al.

1995; Iivari, Hirschheim et al. 1998; Goles and Hirschheim 2000; Iivari, Hirschheim et al. 2000) has also brought attention to other orientations. Particularly the paradigmatic framework of Iivari, Hirschheim et al. (1998; 2000) is an analytical instrument that enables a convenient arrangement of ISD methodologies into several ISD approaches. One of these approaches is the Language/Action Perspective (LAP), which focuses on the use of language to achieve agreement and mutual understanding (Weigand 2003). The LAP approach heavily draws upon the speech-act theory of Austin and Searle, and upon the communicative theory of Habermas. Within both these theories, social beings achieve changes in the (object) world by means of communication. LAP therefore claims that it offers a solution for the mismatch between social perspectives and technical perspectives within ISD.

Several research programs incorporate LAP, of which the DEMO methodology[1] is an example. This methodology has its roots in the SMARTIE project[2] (Dietz 1990a; 1990b; 1991) and was first presented in 1992 (Dietz 1992a; 1992b). Although the methodology has been applied successfully within various practical settings (e.g. Van der Rijst and Dietz 1993; Van Reijswoud, Mulder et al. 1999), neither DEMO nor LAP are in the mainstream of Information Systems Development (Lyytinen 2004). It is unclear if this limited presence is due to shortcomings in LAP theory itself. As ISD is an applied science, a critical analysis from both a theoretical point of view and a practitioner's point of view is required. Although some comments about the applicability of LAP for ISD are available (De Michelis and Grasso 1994; Suchman 1994; Winograd 1994; Bannon, Agre et al. 1995; Ljungberg and Holm 1996), a structured evaluation is not. As of yet, research that draws upon the experiences of the various methodologies incorporating LAP is very rare (Kethers and Schoop 2000 being a notable exception). Therefore, the relationship between LAP theory and practice remains unclear.

## 1.2   Research Questions

This paper aims to devise several recommendations on how the Language/Action Perspective (LAP) can improve its footprint in the community of ISD practice. To understand LAP itself, a critical analysis of LAP's concepts is necessary. To structure this analysis, primary focus resides on the claim that LAP can unify the apparent incompatible social and technical perspectives present in Information Systems Development (ISD). To base the analysis upon both LAP theory and practice, this research includes an assessment of practitioners about LAP as well. Since LAP is a theory and not a methodology that professionals can apply directly, DEMO is chosen as a case example. In order to transfer some findings from the professional's reflection to LAP, DEMO is split into a level of theory and a level of methodology. After identification of the theory and proposed application of DEMO, the relationship with DEMO practice can be established. A comparison between LAP theory and DEMO theory facilitates a critical analysis of LAP. In its turn, this analysis yields the answer whether LAP can undo the apparent dichotomy. This analysis will bring upon several recommendations to improve the practical applicability of LAP's concepts and methodologies. The following derived research questions support the aim of this research:
   1. What is the relationship between DEMO theory and its intended areas of application?
   2. How does the professional application of DEMO differ from its intended application?
   3. Can LAP unify the apparent incompatible social and technical perspectives present in ISD practice?

## 1.3   Research Method

To fully understand a methodology it is useful to analyze the underlying theory that shapes the development of that methodology. In turn, such a theory usually adheres to particular philosophical assumptions. The paradigmatic framework of Iivari, Hirschheim et al. (1998; 2000) incorporates the idea of separating these concepts in order to analyze and categorize ISD methodologies. The framework has a four-tiered structure, consisting of 1) paradigm 2) ISD approach 3) ISD methodology and 4) tools and techniques. Within this framework, a paradigm corresponds with a particular set of philosophical assumptions, separated into the dimensions of ontology, epistemology, research methodology, and ethics. Although from an analytical perspective an ISD approach (ISDA) is an abstract of shared theories and concepts of several similar ISD methodologies, it effectively reflects a research community. Tools and techniques can be seen as instruments to support the actual practice of the ISD methodology (ISDM). This paradigmatic framework is used to analyze DEMO for two reasons. Firstly, it provides an analytical framework to study the underlying philosophical and theoretical assumptions of DEMO as an ISDM. And secondly, it defines the existence of a relationship between DEMO and LAP. This allows that some reflections upon DEMO theory and its practice can be transferred to the entire LAP research community.

---

[1] DEMO is an acronym that has had several different meanings during the last decade. Currently, DEMO is an abbreviation of Design & Engineering Methodology for Organizations (e.g. Dietz and Habing 2004a).
[2] SMARTIE is an abbreviation of *Specification, Modeling, Analysis and Refinement Techniques in Information Systems Engineering*.

Analysis of DEMO by means of the paradigmatic framework mentioned above is purely theoretical, since it does not address the real application of the methodology by practitioners. To study this pragmatic aspect, the experience and judgment of professionals is essential. This requires another research method. To structure the assessment of DEMO by these professionals, the field of DEMO practice is separated into two aspects. The first aspect concerns the area of application. A survey among people who are acquainted with DEMO verifies whether the actual areas of application correspond with the identified areas of DEMO research. Additionally, this survey gives insight into the backgrounds of individual respondents. The second aspect focuses on the impact of DEMO on project methodologies within these areas. Three levels of application represent this impact, being formal application, combined application, and informal application. Together these two aspects structure the project experiences of practitioners. The applied enquiry is part of a workshop, whose attendants are selected based upon their response to the survey. Based upon a discussion about this combined experience, the group of experts gives recommendations for the DEMO research program.

To reflect upon both DEMO theory and practice, a combination of perspectives is required. The multi-dimensional world of Habermas describes these perspectives, distinguishing between the material world, the social world, and the personal world. Mingers and Brocklesby have incorporated these dimensions into a framework that analyzes methodologies for the purpose of combining them in what they call multi-methodologies (Mingers and Brocklesby 1997; Mingers 2000). Another feature of their framework is the notion of different types of activity that need to be undertaken, being appreciation, analysis, assessment, and action. The value of this framework lies in that it allows a combination of the research findings of the applied paradigmatic framework and the expert discussion. Whereas the former results describe the global boundaries of DEMO with respect to the three world dimensions, the latter results express the qualitative appreciation of the different types of activity. Therefore the multi-methodology framework is used to structure the reflection of DEMO.

### 1.4 Outline of This Paper

The questions discussed in §1.2 impose the structure of this paper. Respectively section 2, 3, and 4 address these three questions. Section 5 gives some directions for further research to improve LAP's footprint in the community of ISD practitioners. The first section is an introduction to the background of the research (§1.1), and describes the related question (§1.2). The research method to obtain the answers to these questions is based upon a combination of two frameworks and empirical research (§1.3). These answers form the outline of this paper (§1.4).

Section 2 reviews and analyzes DEMO theory, an example research program that belongs to the LAP research community. First the major concepts of this theory are elaborated (§2.1), derived from a review of key DEMO literature. Next, the identified areas of research that are part of this research program are discussed (§2.2). Both aspects aid the paradigmatic analysis of DEMO theory, which reveals the philosophical assumptions that drive DEMO methodology (§2.3). Based upon this analysis the relationship between DEMO theory and its intended areas of application is identified (§2.4).

Section 3 deals with the application of DEMO from a practitioner's point of view. To link the previous discussion of DEMO theory with the practice of professionals, first DEMO methodology is introduced (§3.1). To get insight into the background of these professionals and the actual areas of DEMO application, a survey is discussed and analyzed (§3.2). A qualitative appreciation of DEMO methodology is part of a workshop, which is structured according to the survey's results (§3.3). The conclusion answers the question how the professional application of DEMO differs from its intended application (§3.4).

Section 4 reviews the Language/Action Perspective when applied to ISD. Being an example methodology, first DEMO is reflected in a combined theoretical analysis and practical assessment (§4.1). Next, the implications of this reflection on the entire LAP community are determined (§4.2). These implications are combined with existing evaluations of LAP to aid a critical evaluation of LAP (§4.3). The conclusion determines if LAP is able to unify the dichotomy present in ISD (§4.4). Finally, section 5 gives several recommendations for further LAP research.

## 2   The Development of DEMO Theory

To understand a methodology, knowledge of its fundamental concepts and their intended application is indispensable. The paradigmatic framework applied in this section disentangles the philosophical assumptions that drive these fundamental concepts, in order to understand the relationship between theory and intended areas of application of DEMO. The first subsection discusses the key concepts of DEMO, followed by an elaboration of the major intended areas of application for the methodology. Based on both these aspects the philosophical background of DEMO theory will be inferred.

### 2.1   Concepts within DEMO Theory

To analyze the theoretical development of the fundamental concepts in DEMO it is necessary to review the key publications of the methodology and its primary sources of inspiration. Although the number of publications about DEMO has been quite stable throughout the past decade, not all publications show a significant contribution to the understanding or development of DEMO. The studied publications are selected by Jan Dietz, the main developer of DEMO and head of the research program that embodies the methodology. It is believed that the publications in Table 2-1 reflect the key developments within DEMO. Due to the review process of most journals, there is a noteworthy delay between acceptance and publication of articles. Therefore the mentioned dates within the remaining of section 2 do not necessarily reflect the actual initiation of certain developments.

| Year | Publication | Publication Type |
|---|---|---|
| 1990 | (Dietz 1990a) | Conference proceedings |
|  | (Dietz 1990b) | Inaugural lecture |
| 1991 | (Dietz 1991) | Conference proceedings |
|  | (Dietz and Widdershoven 1991) | Conference proceedings |
| 1992 | (Dietz 1992a) | Book |
|  | (Dietz 1992b) | Journal article |
| 1993 |  |  |
| 1994 | (Dietz 1994) | Journal article |
| 1995 |  |  |
| 1996 | (Dietz and Mulder 1996) | Conference proceedings |
|  | (Dietz 1996a) | Book |
|  | (Dietz 1996b) | Inaugural lecture |
|  | (Dietz 1996c) | Book section |
| 1997 |  |  |
| 1998 |  |  |
| 1999 | (Dietz and Barjis 1999) | Conference proceedings |
|  | (Dietz 1999) | Conference proceedings |
|  | (Van Reijswoud, Mulder et al. 1999) | Journal article |
|  | (Van Reijswoud and Dietz 1999) | Conference proceedings |
| 2000 | (Dietz and Schouten 2000) | Conference proceedings |
| 2001 | (Dietz 2001) | Journal article |
| 2002 |  |  |
| 2003 | (Dietz 2003a) | Journal article |
|  | (Dietz 2003c) | Journal article |
|  | (Shishkov and Dietz 2003) | Conference proceedings |
| 2004 | (Dietz and Habing 2004a) | Conference proceedings |
|  | (Dietz and Halpin 2004) | Book section |

Table 2-1: Key publications of DEMO

As mentioned in §1.1, the inception of DEMO took place in 1992, when prior work to the conceptual modeling of Information Systems was transferred to that of organization as a social system. The development of DEMO is an advancement of the SMARTIE project, of which the overall goal was to 'search for sound theoretical foundations for the discipline of Information Systems Engineering' (Dietz 1991). The occasion of the DEMO research program is the low success rates of Information Systems. The identified cause is the lack of models capable of analyzing the essential aspects of organization. Often this results in leaving aside the entire analysis of organization during the inception of Information Systems Development. Therefore, a shortage of precision and details when defining the context of Information Systems is eminent (Dietz 1992a). DEMO defines three key concepts interlinked in a theory to overcome this shortage. Firstly, organization is formally defined as a social system with a finite number of elements that collectively exhibit particular behavior. Secondly, the only active elements of such a social system are human beings, who operate on and communicate about things in the object world. And thirdly, the communication of these human beings has three different aspects, i.e. essential, informational, and documental. Each of these concepts is elaborated hereafter.

DEMO's definition of a system draws upon Bunge (1979), who among other things clearly distinguishes between an assembly of elements and an aggregate of elements. The adopted universe of Bunge is a system with only one instance. To study its subsystems, application of the set of theories that focus on the structural characteristics of systems known as systemics[3] is required (Bunge 1979: xiii, 1). Such a formal approach to the study of systems is similar to the practice of ontologists. Therefore the organization as system can be regarded as an ontological concept. The precise definition of an organization according to DEMO is as follows (Dietz 2001).
 Something is an organization if and only if it fulfills the next properties:
- It has composition, i.e., it is composed of actors, where an actor is defined as one or more subjects in a particular role. These actors act on the basis of assigned authority and with corresponding responsibility.
- It has structure, i.e., the actors influence each other. Two kinds of mutual influencing are distinguished. Interaction consists of executing transactions. Interstriction consists of taking into account the results or the status of other transactions when carrying through a transaction.
- It has boundary. The composition (i.e., the set of constituting actors) is divided into two subsets, called the kernel and the environment, such that every actor in the environment influences, either through interaction or through interstriction, one or more actors in the kernel, and such that there are no 'isolated' parts in the kernel. The closed line that separates the kernel from the environment is called the boundary.

A subject-object dichotomy drives the conceptual model of the organization as system. This model includes both the shared social world – regarded as a intersubject system – of subjects and the object world these subjects act upon. By analogy of Wittgenstein, a subject's[4] knowledge about the object world consists of facts (Dietz 1992b). But subjects also actively change the state of the object world as their actions lead to new facts. To stress that subjects require an occasion before they act, DEMO discriminates between state facts (stata – things known) and agenda facts (agenda – things to do). In accordance with Searle's principle of expressibility, DEMO states that 'every elementary fact can be expressed by an elementary sentence in natural language' (Dietz 1992b: 229). The communicative theory of Habermas – a refinement of the speech-act theory of Austin and Searle (Dietz and Widdershoven 1991) – provides a framework to analyze the exchange of elementary sentences between subjects. Figure 2-1 displays the communicative framework and its interpretation within DEMO. As the example conversation of ordering a beer suggests, successful (factagenic) performative conversations lead to new stata[5] – as a result of subjects who are committed to agenda. On the other hand, informative conversations only reproduce known facts. Habermas differentiates between three different claims[6], of which the dominant claim within a class of conversation is displayed. If these claims are not questioned by one of the subjects, the conversation will be successful. In reality, most communication between subjects is not explicit. But DEMO does provide a complete framework for the study of transactions (e.g. Dietz 2003c), such that these implicit aspects can be recognized despite of ambiguity within the context.

---

[3] Systemics is rooted in the unified approach of the General Systems Theory, as advanced by Von Bertalanffy (1950) and Boulding (1956), among others (Bunge 1979: 1).

[4] The word actor is a role concept in DEMO to facilitate a model of grouped subjects. However, within the remaining of this section the word subject is maintained to stress the philosophical roots of this concept.

[5] Stata therefore belong to the intersubject system and not to the object world. This implies that facts in the object world do not necessarily have to correspond with their subjective counterpart. A clear example is an umpire who claims a ball to be out in a tennis match, while actually the ball was in (Dietz 1991: 452).

|  | performative conversation | | | | informative conversation | | |
|---|---|---|---|---|---|---|---|
|  | actagenic | | factagenic | | | | |
| DEMO / Habermas | Directives | Commissives | Statutives | Acceptives | Interrogatives | Assertives | |
| **Constativa** |  |  |  |  | Does it rain? | It rains | *Claim to truth* |
| **Regulativa** | I'd like to have a beer | One moment please | Here you are | Thanks |  |  | *Claim to justice* |
| **Expressiva** |  |  |  |  |  |  | *Claim to sincerity* |

**Figure 2-1: Communicative theory of DEMO (adapted from Dietz and Schouten 2000)**

While the communicative theory focuses on the essential aspects of communication between subjects, and the organization as system formally defines the relationships between subjects within a specific boundary, these concepts combined still only provide an abstract, high-level blueprint of organization. Issues such as supportive technology and the assignment of roles are not mentioned, while they are required to understand the realization of the essential aspect of organization from the blueprint. To facilitate part of this requirement, DEMO incorporates a layered approach that is adapted from semiotics. Based on the observation that there is no information without communication, the notion of information is closely related to communication (Dietz 1999). Stamper's semiotic ladder distinguishes a physical, syntactic, semantic, and pragmatic aspect of information. These aspects can be put in a sequence of three abstraction levels, known in DEMO as the documental, informational, and essential model (Dietz 1994). The physical aspect is ignored, since it is not an abstraction. The essential model is now equal to the previously elaborated model of organization, and is shaped by performative conversations. The informational model deals with informative conversations, since they only reproduce known facts and do not change the state of the object world. Finally, documents abstract from the physical carriers of information that support the informational model. Therefore, the layered approach in DEMO is an interlinked representation of various abstraction levels of organization that allows for a separation of concerns during organizational analysis.

## 2.2   Identified Areas of Research

The theory of DEMO has been applied to various areas of application during the last twelve years. Soon after the observation that new methodologies for the support of conceptual modeling of Information Systems were required, the major focus of DEMO has been analysis of organization. While at first analysis focused upon supporting ISD, a few years later the DEMO research program directed its efforts at other areas of research. Development of DEMO was parallel to that of Business Process Reengineering (BPR) (e.g. Davenport and

---

[6] The 'claim to power' which was part of Habermas' original theory is abandoned, as it was regarded as a variation of 'claim to justice' later on by Habermas himself (Dietz 2001).

Short 1990; Hammer 1990; Keen 1991; Scott Morton 1991; Davenport 1993; Hammer and Champy 1993), which resulted in several publications about Business Process Redesign (e.g. Dietz 1994; Dietz and Mulder 1996). When the management hype of BPR faded away in the second half of the 1990s, DEMO research shifted towards organization engineering. In summary, the three global areas of research are I – ISD, II – Business Process Redesign, and III – Organization Engineering. Table 2-2 shows the global development and areas of research of the DEMO research program. The type of publication refers to the classified area of research. While a certain chronological development of the research focus can be identified, each of the addressed topics is recurrent. Therefore each of the three topics is elaborated separately within the remaining of this subsection.

| Year | Type | Research topic | Citation |
|---|---|---|---|
| 1990 | I | Conceptual modeling of information systems | (Dietz 1990a; 1990b) |
| 1991 | I | Conceptual modeling of social systems | (Dietz 1991) |
|  | I | Refinement of speech act theory | (Dietz and Widdershoven 1991) |
| 1992 | I | Conceptual modeling of social systems | (Dietz 1992a; 1992b) |
| 1993 |  |  |  |
| 1994 | II | Methodology for Business Process Redesign | (Dietz 1994) |
| 1995 |  |  |  |
| 1996 | II | Reengineering organizations through IT | (Dietz and Mulder 1996) |
|  | II | Business Systems Engineering | (Dietz 1996b; 1996a) |
| 1997 |  |  |  |
| 1998 |  |  |  |
| 1999 | II | Simulation of Business Processes | (Dietz and Barjis 1999) |
|  | II | Business Process Modeling | (Dietz 1999) |
|  | I | Information Systems Development approach | (Van Reijswoud, Mulder et al. 1999) |
|  | II | Methodology for Business Process Redesign | (Van Reijswoud and Dietz 1999) |
| 2000 | III | Support of virtual organizations | (Dietz and Schouten 2000) |
| 2001 | III | Organization Engineering | (Dietz 2001) |
| 2002 |  |  |  |
| 2003 | III | Reference framework for business processes | (Dietz 2003a) |
|  | III | Generic patterns in modeling processes | (Dietz 2003c) |
|  | I | Derivation of use cases from business processes | (Shishkov and Dietz 2003) |
| 2004 | III | Reference ontology for a class of organizations | (Dietz and Habing 2004a) |
|  | I | Synthesis of DEMO with Object-Role Modeling | (Dietz and Halpin 2004) |

**Table 2-2: Overview of general areas of research**

As mentioned in §2.1, DEMO pinpoints the lack of formal organizational analysis during ISD as the main cause of failure for unsuccessful IT projects. Perhaps influenced by constructivist background of the field of informatics, this analysis is due to be *formal* in order to obtain an objective blueprint of the organizational context of Information Systems. This implies that subjective observations about organization are excluded from the analysis, since they are highly open to discussion between different analysts. The relationship between the high level model of organization and supportive Information Systems is facilitated within DEMO by various interlinked models[7]. As explained in the previous subsection, communication between subjects can lead to new actions that lead to new stata. And actors will perform these actions, because they are committed to their agenda. DEMO defines interdependent relationships and conditions between actions – as part of transactions – in a model of business processes. The related information is modeled in a state model. Therefore, the inception of ISD according to DEMO is a process that selects appropriate parts of the organizational model in order to redefine or support its implementation in terms of Information Systems. DEMO provides a starting point for the definition of functional requirements of such Information Systems (Shishkov and Dietz 2003), and thereby positions itself at the requirements engineering phase of ISD. As a result, ISD only changes the implementation of the essential business processes and not their definition.

While organizational analysis according to DEMO was quite modest in its aspirations during the first years of the 1990s, the uprising of Business Process Reengineering gave birth to the realization that DEMO analysis could also be used to aid organizational change. The first major article appeared in 1994, which aimed to 'develop an original contribution from the discipline of informatics to BPR methodologies' (Dietz 1994: 233). During ISD, the essential model of organization was primarily used to understand the context of Information Systems that reside at the informational level. The application of DEMO for Business Process Redesign focuses primarily at the essential level – the domain of Business Systems. The theory has been refined in the following years, leading to an explicit methodology for the purpose of Business Process Redesign in 1999 (Van Reijswoud and Dietz 1999). To analyze the impact of proposed changes to the business processes of an organization, simulation of business processes has been an important area of research (Dietz and Barjis 1999). However, it is worth mentioning that despite several changes in application of DEMO methodology, the underlying concepts of DEMO theory as applied in ISD remained quite stable.

---

[7] Section 3.1 discusses the various models and their relationships in more detail.

Somewhere in the second half of the 1990s, focus of DEMO shifted from Business Process Redesign to Organization Engineering. This coincided with the diminishment of Business Process Reengineering as management hype (Davenport 1995). The observed gap between professionals from organization sciences and professionals from information processing sciences became a new focal point of research in 2001 (Dietz 2001). The notion of this gap, however, was mentioned earlier: it was present as early as 1996 (e.g. Dietz 1996b). It is therefore important to determine whether this focus represents a substantial or a cosmetic change. To bridge the mentioned gap, DEMO considers knowledge of the construction and operation of business processes as a requirement. The white-box model associated with this approach is a contrast with the popular black-box – or teleological – models as found in organization sciences. While the former illustrates how an organization is constructed by business processes, the latter is mostly a functional representation of the organizational operation. DEMO provides a constructional view of organization as it focuses on the subjects within organization and on the coordination of their actions. As such, DEMO does not concern itself with an organization's mission, but only with the means of realizing it. In addition, preliminary research into the subject of role assignments indicates that DEMO can be linked with fields as Human Resource Management as well (Dietz 2003a).

### 2.3   Paradigmatic Analysis of DEMO Theory

The previous two subsections elaborated the key concepts of DEMO theory and the intended areas of application. Within the paradigmatic framework of Iivari, Hirschheim et al. (1998; 2000) these key concepts are grouped into an ISD approach (ISDA). DEMO belongs to the ISDA of the Language/Action Perspective (LAP). The intended areas of application are an indication of the philosophical assumptions that drive the ISDA. As explained in §1.3, the paradigmatic framework is applied for two reasons. This subsection focuses on the first, being a structured analysis of DEMO theory in order to understand the relationship between DEMO theory and its intended areas of application. §4.2 will compare this analysis of DEMO with the analysis of the LAP as performed by Iivari, Hirschheim et al. (1998). The framework and its concepts are displayed in Table 2-3. Because this analysis focuses on the philosophical assumptions underlying DEMO theory, this paradigmatic analysis takes place exclusively on the level of ISD Paradigm. Unlike the original analysis of LAP that was solely based on text interpretation, personal interviews with the main developer of DEMO both augment and discuss this analysis. Each of the related aspects is elaborated in the remainder of this subsection.

| Level of abstraction | Aspects | Interpretation |
|---|---|---|
| *ISD Paradigm* | | A set of philosophical assumptions and beliefs underlying every ISDA and ISD that allows ISDAs to be grouped into a number of paradigmatic positions. |
| | Ontology | What is assumed to be the nature of IS. |
| | Epistemology | What human knowledge is and how it can be acquired. |
| | Methodology | Preferred research methods for continuing the improvement of the ISDA as well as how the ISDA was developed and justified in the first place. |
| | Ethics | The values that ought to guide IS research. |
| *ISD Approach* | | A set of related features that drive interpretations and actions in ISD. |
| | Goals | General purpose of the ISDA. |
| | Guiding principles | The common "philosophy" of the ISDA, which ensures that its ISDM instances form coherent |

|  |  | holes. |
|---|---|---|
|  | Fundamental concepts | Focus and unit of analysis in ISD. |
|  | Principles of the ISD process | Essential aspects of the ISD process in the ISDA. |
| *ISD Methodology* |  | A codified set of goal-oriented 'procedures' which are intended to guide the work and cooperation of various parties (stakeholders) involved in the building of an IS application. |
|  | Relationships between techniques | N/A. (*Definition not presented in framework.*) |
|  | Detailed ISD process | N/A. (*id.*) |
| *ISD Technique* |  | A well-defined sequence of elementary operations that more or less guarantee the achievement of certain outcomes if executed correctly. |
|  | Detailed concepts | N/A. (*Definition not presented in framework.*) |
|  | Notations | N/A. (*id.*) |

**Table 2-3: Four-tiered conceptual structure of the paradigmatic framework (adapted from Iivari, Hirschheim et al. 1998; 2000)**

Within the dimension of ontology, Iivari, Hirschheim et al. (1998: 172) identified five phenomena within IS research, which are maintained in this analysis to facilitate a comparative evaluation. These phenomena are 1) Data/information, 2) Information Systems, 3) Human beings, 4) Technology, and 5) Organizations and society. In general, DEMO adheres to the world view of Bunge, who views the world as a system with only one instance[8]. In more detail, DEMO has a remarkable view on *data and information*, in that it regards them as derived from acts that result in new facts. The class of regulativa is therefore constitutive. Nevertheless, the class of constativa is descriptive in that it describes the world as observed by subjects. A fact – being an agendum or a statum – is essential, while information and data are on two lower levels of abstraction. DEMO theory regards *Information Systems* as social systems, albeit on a supportive level of Business Systems that shape the essential level of organization. This implies that – either in part or even in its entirety – the Information System may be implemented by technology, because on the essential level human beings are still responsible for the results and actions of these systems[9]. With respect to *human beings*, DEMO adheres to voluntarism. As a result of the three validity claims of Habermas, and the commitment of subjects to agenda, human beings have free will, although they are bound to the larger organizational context. *Technology* in DEMO is supportive in the way that it implements the essential aspect of organization. Since human beings are ultimately responsible, technology is subject to human choice. Finally, DEMO considers o*rganizations and society* predominantly from a structural point of view. Although role assignment and the distribution of work may be questioned at any time, DEMO regards these aspects as implementation issues, whereas the underlying structure is assumed to be stable.

The *epistemology* of DEMO is predominantly positivist. Although subjects can question the illocution and proposition of their conversation partner, ultimately the exposed behavior is bound to socionomic laws (Dietz 2003c). Because the class of expressiva is excluded from DEMO theory, psychological and emotional states are deemed irrelevant for understanding of the business communication within organization (Dietz and Schouten 2000). Another indication of the positivist epistemology is the absence of the role of the analyst within DEMO, suggesting that the analyst is an objective observer, isolated from the object of study. With respect to *research methodology*, DEMO applies constructive conceptual development as it seeks to produce models and related procedures. Indeed several real-life cases have been reviewed, but their related papers seem to be more proof of concepts of DEMO theory than an applied idiographic[10] method. Finally, regarding *ethics of research* DEMO supports a means-end orientation for the role of IS science. The strongest indication of this proposition can be found in the papers about Business Process Reengineering, where the role of DEMO is limited to redesign, whereas the motives of the reengineering movement and its accompanied organizational change remains unquestioned. With respect to the value of IS Research DEMO seeks to improve organizational effectiveness. This is supported by making organization a transparent system that is understandable for everyone.

---

[8] See §2.1 for an elaboration.
[9] An illustrative example of this view of Information Systems is the discussion of an elevator control system designed as a social system (Dietz 2003b).
[10] Brown (1992: 154-5) defines the idiographic approach as "the thorough study of individual cases, with emphasis on each subject's characteristic traits". It is opposed to nomothetic approaches that can be characterized as "the study of a single variable in many subjects for the purpose of discovering general laws or principles of behavior".

## 2.4   Conclusions

DEMO emerged within the discipline of Information Systems Development as a methodology to analyze the organizational context of Information Systems. Within each of the successive areas of research, i.e. Business Process Redesign and Organization Engineering, DEMO retained its objective of providing a formal methodology to model organization. The underlying approach can be characterized as positivist, constructive conceptual development, as the resulting blueprint of organization is independent of the analyst. This is only possible within a realist ontological position, which DEMO bases upon from the systemic theory of Bunge. Although DEMO does not question its means of application, it easily fits within a technocratic management tradition that regards organization as the means to achieve corporate goals. DEMO's abstraction from human beings to rational actors dismisses any social complications that may arise within organization. Accordingly it primarily regards organization as a problem of coordination.

The relationship between DEMO theory and its intended areas of application is based upon the analysis and construction of formal systems. As such, the real strength of DEMO is its ability to analyze and define complex problem situations. These problems mainly concern structural problems that typically arise in large organizations or complex networks of interdependent organizations. Virtual organizations with many involved parties are a profound example. Although at a first glance the incorporated communicative theory of Habermas might suggest DEMO is a social theory, in reality it only determines the socionomic laws that regulate communication. Nevertheless, DEMO has the remarkable position that human beings are responsible for the working and effects of Information Systems. Actually, DEMO states that ultimately these human beings are responsible for a part of organization's operation, including its supportive systems. This could provide an opening to supplementary methodologies that analyze organization from a more social, interpretivist perspective.

## 3   The Professional Application of DEMO

While the previous section focused on DEMO theory, this section elaborates the practical application of DEMO. Insight into this practical application is based upon a combined quantitative and qualitative research method. As most practitioners are more concerned with models and techniques than underlying theory, the following section first presents a brief introduction to DEMO methodology. Next, a survey and its results are discussed. Finally, the survey results aid the structure of an expert discussion that is part of a workshop, held in a Group Decision Room.

### 3.1   Introduction to DEMO Methodology

Section 2 discussed and analyzed DEMO on the levels of ISD Approach and ISD Paradigm. Both levels are not directly related to the real-life application of practitioners. Although one can assume that the application of a methodology requires at least basic knowledge of the underlying theory, it is likely that knowledge of specific tools and techniques is much more relevant within a goal-oriented context. Within the paradigmatic framework the accompanied levels are known as ISD Methodology and ISD Technique[11]. Therefore, to smooth the progress of communication with professionals the primary focus of this section will reside on these two levels. Because the Level of ISD Approach (ISDA) shapes the level of ISD Methodology (ISDM), Table 3-1 summarizes the ISDA level of DEMO as elaborated in §2.1. To facilitate the comparative evaluation of DEMO and LAP in §4.2, the summary also shows the interpretation of LAP by Iivari, Hirschheim et al. (1998: 168).

---

[11] See §2.3 for an elaboration of both these concepts.

| Aspect | DEMO Interpretation | LAP Interpretation |
|---|---|---|
| Goals | To provide a methodology for modeling business communication aligned with supportive Information Systems and ICT Infrastructure. | To provide a methodology for modeling communicative action in organizations, especially speech acts of changes: creating, maintaining, reporting, modifying and terminating organizational commitments. |
| Guiding principles | Ultimately humans are responsible for action; Information Systems are only supportive to realize social commitments; Organization in essence is a coordination problem. | An information system is a social system only technically implemented; An information system is a communication system (mediating speech acts); ISD is formalization of professional (work) language. |
| Fundamental concepts | Communication; Information; Action; Organization. | Speech act; Illocutionary points; Propositional content; Discourses/conversations. |
| Principles of the ISD process | Discourse analysis[12]; Formal modeling; Separation of concerns by means of different abstraction levels. | Discourse/conversation analysis; Analysis of the propositional content. |

**Table 3-1: Summary of DEMO and LAP ISDAs**

While the concepts within DEMO theory have remained stable throughout the past decade[13], the models and diagrams that belong to DEMO methodology have been subject to many changes. Not only terminology altered frequently, but also insights from example Petri Nets (Dietz and Barjis 1999) and Object Role Modeling (ORM) (Dietz and Halpin 2004) have had major implications on the models and diagrams of DEMO methodology. A full overview of the methodological development of DEMO is outside the scope of this paper. The remainder of this subsection is therefore just an introduction into the latest state of affairs of these models and their relationships. Dietz and Habing (Dietz and Habing 2004b; Dietz 2005) discuss four aspect models, which are 1) Construction Model, 2) Process Model, 3) Information Model, and 4) Action Model. Each of these models resides as the essential level of organization, although they do have an increasing order of detail. The Construction Model represents the blueprint of organization, as it deals with actors and transactions in the organization as system. Both the Process Model and the Information Model support the Construction Model. The Process Model specifies the causal and conditional relationships between the identified transactions of the Construction Model. The Information Model is a conceptual schema of the things and facts that appear to be relevant. Finally, the Action Model serves as a guideline for actors who deal with their agenda, as it specifies the action rules for the various types of agenda.

---

[12] Ljungberg and Holm (1996) describe a discourse as 'a globally managed sequence of communicative actions (speech acts), forming a coherent and predetermined course of action leading to a goal'. The accompanied schematic theory construction of discourse analysis is a contrast with the empirical approach of conversation analysis, which is rooted in ethno methodology.

[13] See §2.1 for an elaboration of the three key concepts of DEMO theory.

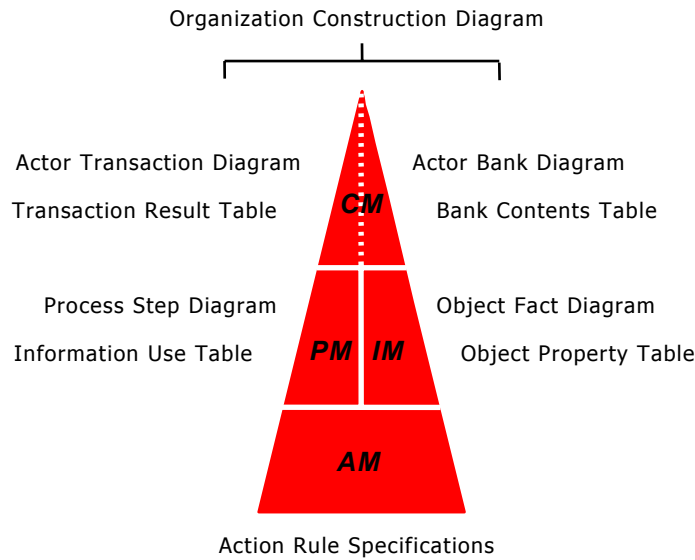

**Figure 3-1: Relationship between DEMO aspect models and diagrams (adapted from Dietz 2005)**

Although DEMO has no strict procedure to define the various models, in practice most projects follow the same general procedure[14]. Several diagrams and tables visualize the aspect models of DEMO methodology. Figure 3-1 depicts the relationships between these visualizations and the aspect models. The Action Model is even represented by pseudo-code that specifies the various action rules. The core diagram that visualizes the Construction Model (CM) is the Actor Transaction Diagram. It displays both the internal (elementary) actors and external (composite) actors, related to each other by transactions. The system boundary depends on the object of analysis, but typically represents the border of an organization or a sub organization. The Process Step Diagram visualizes the Process Model (PM) that defines the relationships between the identified transactions. Each point of initiation denotes a specific business process, which typically consists of several interrelated transactions. As the diagram is a special type of Petri Net, it is suited for process simulation. The Object Fact Diagram is an ORM-like diagram that displays the fact types, event types, and object classes that model the persistent information as part of the Information Model (IM). Lastly, the aforementioned pseudo-code that represents the Action Model (AM) defines the conditional behavior of actors committed to agenda by specific action rules.

## 3.2   Real-life Application of DEMO

A combination of quantitative and qualitative research methods aids the analysis of the application of DEMO by practitioners. A survey gives insight into the background of the research population and the general fields of application, while an expert discussion aids a qualitative assessment of DEMO methodology. The experts are selected based upon their response to the survey. After an initial mailing to 284 persons known to be acquainted with DEMO, 114 valid addresses remained. Each of these persons received an electronic survey, of who 50 gave a usable response (43.86%). Several others indicated that they lacked experience with DEMO, or lost their professional interest in DEMO. Out of these 50 responses, 25 were requested to join the expert discussion – each of them had at least finished one DEMO project. In total 19 persons joined the workshop, including several interested persons who were invited separately. The remaining of this subsection elaborates the background of the research population and identifies the global areas of application. Section 3.3 gives the assessment of DEMO applied in practice.

The survey consisted of sixteen questions, of which only the first seven are relevant for this section. The complete answers to these questions can be found in Appendix 1, as the remainder of this subsection only discusses the highlights. In total four questions of the survey address the background of the respondent. Combined they give an impression of the environment and job activities of DEMO practitioners.

---

[14] See Dietz and Habing (2004b) for an elaboration.

*Q 1.* Do you mostly work for your employer (internal) or for clients (external)?
There is no dominant working environment for the practitioners, as 44% mostly works internal and 56% external.

*Q 2.* To which sectors can your clients or employer be classified?
More than a quarter (27%) of the respondents works in the service industries, or has clients in that sector. The public sector and various sector (including health and building industry) each score 19%. The relatively low share (15%) of automation can be explained by its service orientation, which by default is focused at external clients.

*Q 3.* What is the size of the organizations that commission your work?
Almost a third (31%) of the organizations has more than 1000 employees.

*Q 4.* What amount of time do you allocate for each of the following activities?
34% of the time is spent on consulting and 15% on designing. Implementation has a relatively low share of 10%, which suggests the respondents – with respect to ISD – are more concerned with requirements engineering.

Next to the background of the respondent, the survey includes three questions to define the practical application of DEMO.

*Q 5.* For which areas of application have you used DEMO?
Business Process Redesign has a share of 43% and Information Systems Development a share of 37%. Apparently two out of three of the research areas identified in §2.2 correspond with the global areas of application.

*Q 6.* On what level do you apply DEMO?
Despite the assumption that DEMO has a low impact on the ISD community, remarkably few only apply DEMO on a personal level (19%). Possibly the nature of DEMO projects requires at least several team members to be acquainted with the methodology.

*Q 7.* What is the average scheduled time for DEMO projects?
The average scheduled time for DEMO projects is 4–6 months, with an almost even distribution towards both shorter and longer projects. Long term projects – exceeding 1 year – are exceptional.

Because the total size of the research population is estimated to be 1000 persons[15], the amount of responses of the survey is not significant. Nevertheless, the response does give an impression of the background of DEMO practitioners and their applications. Based upon the nature of DEMO theory as well, it seems the added value of DEMO methodology is its ability to structure complex organizational problems. In particular, these problems surface within the general areas of application of either Information Systems Development (ISD) or Business Process Redesign (BPR). As the main concern of DEMO is coordination between actors, the complexity of these problems is strongly related to the amount of actors and transactions. This is typically so in large commissioning organizations or virtual organizations consisting of many different parties. Although DEMO does not discriminate between types of organizations, in practice it is more applied to organizations with immaterial processes, such as those present in the service industries. Compared to the identified areas of research in §2.2, actual application of DEMO is limited to the first two of these areas: ISD and BPR.

### 3.3 The Assessment of DEMO by Practitioners

As mentioned in the previous subsection, 19 practitioners were willing and able to join an expert discussion. This discussion was part of a workshop that lasted four hours in total. To structure the assessment of DEMO and to devise several recommendations for future DEMO developments and activities, the first part of the workshop focused on listing DEMO projects and experiences. After a break, several highlights were selected for a structured discussion. A Group Decision Room facilitated each of these topics. Such an electronic meeting application has as advantage that it allows every attendant equal opportunity to contribute to the discussion. Especially in meetings consisting of many participants, discussion is likely to be dominated by a few attendants otherwise. The overall setup of the workshop was discussed by two small groups – consisting of a maximum of five people – beforehand. This ensured the subsequent agenda of the workshop was in line with the participant's expectations about the workshop. The remainder of this subsection refines the previously identified areas of application. In addition, the recommendations about the future of DEMO are elaborated. Section 4.1 will address the scope of DEMO organizational analysis, which was another part of the workshop.

---

[15] The DEMO Center of Expertise claims to have familiarized about 700 professionals with DEMO. Taking into account these professionals may have introduced DEMO to their acquaintances, a number of 1000 persons familiar with DEMO seems fair.

The survey indicates DEMO's major areas of application are Information Systems Development (ISD) and Business Process Redesign (BPR). Combined with a breakdown in abstraction levels about DEMO's application (formal, combined, and informal) these aspects provide a structure to list project experiences. Each of the participants was asked to enter his particular project experiences into one of the resulting six categories. Projects that encompassed both BPR and ISD were categorized into the former area. In total 67 projects were identified, of which 44 concerned BPR and 23 concerned ISD. For both areas of application, DEMO is mostly applied in combination with other methods and techniques. UML and Petri Net are techniques that particularly came forward. In general, the most popular model of DEMO methodology is the Construction Model, for both ISD and BPR. The Process Model is typically used within the field of BPR, while the Information Model is more applied within ISD. The Action Model is not part of any project experience. Table 3-2 displays the exact numbers associated with each category. The score indicates the expressed interest of the workshop's participants into future research and development. The participants stress that more research is necessary to provide the required interfaces and procedures that are part of a combined project methodology.

| Application | Number | Score | Variability |
|---|---|---|---|
| 1.  ISD (combined) | 14 | 7.1 | 50% |
| 2.  BPR (combined) | 27 | 7.0 | 46% |
| 3.  BPR (informal) | 6 | 5.7 | 65% |
| 4.  BPR (formal) | 11 | 5.2 | 58% |
| 5.  ISD (formal) | 2 | 4.8 | 57% |
| 6.  ISD (informal) | 7 | 4.6 | 61% |

**Table 3-2: Desirability of research focus**

Although section 3.1 stated that most DEMO projects follow the same general procedure, the experience of the workshop's participants indicates the actual application of DEMO is more diffused. In practice, DEMO diagrams are combined with other methods and techniques. In addition, the selection of the accompanied DEMO models is based upon a fit-for-use criterion. It is notable that none of the participants expressed particular interest into ideal-type project methodologies. Rather they are more interested in theoretically sound combinations with de facto standards, particularly so within the field of ISD. The listed projects show a wide range in scope, ranging from the devising of Enterprise Architectures to the designing of software tools. This indicates DEMO's concepts are appealing for many different scenarios and settings. As such, these concepts in general are not questioned. Most concerns about DEMO revolve around the level of DEMO methodology. To improve the applicability of DEMO, professionals express the need of an active community of practitioners. Additionally, DEMO research should provide directions for possible combinations with other methods and techniques.

### 3.4   Conclusions

Out of the three identified areas of research, only Information Systems Development and Business Process Redesign are applied by DEMO practitioners. Organization Engineering therefore seems more of a theoretical concept than a practical one. Although DEMO methodology offers an integrated design approach, in practice most professionals use aspects of DEMO as they see fit. Particularly within the field of ISD many de facto standards compromise full application of DEMO methodology. But as the concepts of DEMO theory remain appealing for projects of different scope and complexity, practitioners seek combinations and interfaces between DEMO methodology and other methods and techniques. A lack of theoretical backing about these combinations currently hinders the application of DEMO. Additionally, as several research papers do address some of these combinations, a lack of knowledge about these developments is a factor that contributes to the obstruction of DEMO application. Therefore, the problem lies in both the quantity and the distribution of information.

As the concepts within DEMO theory remain unquestioned by practitioners, the theory seems able to meet the requirement of providing sound backing for parts of ISD practice. As such, the professional application of DEMO primarily differs on the level of methodology with the intended application. Especially for de facto standards that lack formal semantics, DEMO offers an appealing, augmentative theory. Rather than to replace existing de facto diagrams and techniques, DEMO should supplement the accompanied syntax with its own semantics and pragmatics. Although this introduces the risk that these diagrams are misinterpreted by people unfamiliar with DEMO theory, the potential for meeting a real practitioner's need might outweigh this drawback. Within this respect, full elaboration of DEMO as a reference methodology can remain valuable for educational and scientific purposes. But as the need for interfaces and combination between DEMO methodology and current standards indicates, replacement of de facto diagrams and techniques is likely a fallacy.

## 4 Challenges of the Language/Action Perspective

To determine if the Language/Action Perspective is able to unify the dichotomy present in ISD, LAP's concepts and assumptions are critically analyzed. First, DEMO as an example methodology of LAP is reflected from both a theoretical and practical point of view. The next subsection compares DEMO theory with LAP to determine which parts of DEMO's reflection can be transferred. Finally, this reflection is augmented by comments from other research papers about LAP.

### 4.1 Reflection of DEMO Theory and Practice

Organizational analysis is the key focal point of DEMO. Section 2.2 identified three major areas of research, whereas chronologically the first was Information Systems Development (ISD) – in particular the requirements engineering phase. The later research topics all build upon this prior research. Because organizational analysis is such a wide and diffuse concept, it is still relatively meaningless if DEMO is to be understood on a methodological level. Section 2 concluded that organization according to DEMO is primarily a problem of coordination. This section further elaborates upon this conclusion with use of the multi-methodology framework by Mingers and Brocklesby (Mingers and Brocklesby 1997; Mingers 2000). The framework partitions methodologies into twelve distinct facets and indicates the accompanied applicability of the methodology under review. Although this framework originated within the context of Operational Research/Management Science and DEMO belongs to the field of ISD, both have in common that they focus on organization as the object of study. Therefore the multi-methodology framework structures this high-level reflection of DEMO theory.

As §2.1 mentioned, DEMO theory is based upon the three world dimensions of Habermas. The class of constativa belongs to the material (object) world, the class of regulativa to the social (intersubject) world, and the class of expressiva to the personal (subject) world. The framework of Mingers and Brocklesby is based upon the same three dimensions. Although the material world exists independently from human beings, these same human beings do observe and change this world shaped by their own experiences and ideas. The personal world is only accessible by the corresponding individual to whom it belongs, as it is 'the world of our own individual thoughts, emotions, experiences and beliefs' (Mingers and Brocklesby 1997: 493). Finally, the social world is formed by a complex of relations between several individuals, who among other things share language and meaning. DEMO excludes the personal world from its theory, since the associated class of expressiva is regarded to be irrelevant for understanding of the business communication of organization[16]. The subject-object dichotomy that drives the conceptual model of organization in DEMO therefore corresponds respectively with the social world and the material world of the multi-methodology framework.

Besides the aforementioned world dimensions, the multi-methodology framework also discusses four different types of activity. Each of these types of activity is expressed by a qualitative value of the practical application of DEMO methodology. To obtain a basis for these values – rather than to present the opinion of just one individual, participants of the workshop[17] were asked for their opinion. They did so by rating several propositions on a scale of 1 (low appreciation) to 10 (high appreciation). Table 4-1 displays the analysis of DEMO by means of the multi-methodology framework. A dark grey color indicates a high appreciation and a light grey color denotes a low appreciation. Each cell contains the interpretation of Mingers and Brocklesby as well. Appendix 2 gives an overview of the actual qualitative values. As the personal world is not part of DEMO theory, the qualitative values of the four accompanied types of activities are all zero.

---

[16] See §2.3 for an elaboration.
[17] The setup and results of the workshop are discussed in section 3.

|  | **Appreciation of** | **Analysis of** | **Assessment of** | **Action to** |
| --- | --- | --- | --- | --- |
| **Social** | Social practices, Power relations | Distortions, conflicts, interests | Ways of altering existing structures | Generate power and enlightenment |
| **Personal** | Individual beliefs, meanings, emotions | Differing perceptions and personal rationality | Alternative conceptualizations and constructions | Generate accommodation and consensus |
| **Material** | Physical circumstances | Underlying causal structure | Alternative physical and structural arrangements | Select and implement best alternatives |

**Table 4-1: Analysis of DEMO by means of the multi-methodology framework**

Although DEMO theory is based upon the same world dimensions of the multi-methodology framework, it adheres to a more narrow interpretation. Because the practitioners were all familiar with DEMO's version, the provided explanation and context – being a more general elaboration of the three dimensions – were difficult to grasp for some. Possibly this affected their interpretation of the offered propositions, which were used to obtain their opinion about the different types of activity. Nevertheless, the valued characteristics of DEMO methodology are in line with expectations. Because DEMO abstracts from human beings to actors, the methodology deals more with causal relationships between processes and activities. Typically role assignments, which are a realization issue, are excluded from DEMO's analysis. The practitioners did agree however, that this concept of abstraction is quite useful to enlighten participants about current arrangements. The combined reflection of DEMO theory and practice indicates DEMO has a definite scope with respect to organizational analysis.

### 4.2   Implications of DEMO Reflection upon LAP

Although DEMO belongs to the LAP research community and therefore shares at least some concepts and viewpoints, the previous sections identified several major aspects of divergence as well. At a methodological level such differences are not surprising, because models, diagrams, and procedures reside at such a low level of abstraction that they are highly subject to personal choice and liking. Nevertheless, section 3.1 also shows dissimilarity on the level of ISD Approach. The reflection of DEMO theory and practice as elaborated in the previous subsection can therefore not immediately be transferred to LAP. To decide which observations are valid for LAP, first a comparative evaluation of DEMO theory and LAP is required. To explain the identified differences and similarities, a comparison of the paradigmatic assumptions between both theories is included as well. Based on this analysis, the implication of DEMO's reflection upon LAP is elaborated.

Section 3.1 gave an overview of DEMO and LAP on the level of ISD Approach. Table 4-2 gives an overview of the paradigmatic assumptions of both DEMO theory and LAP (adapted from Iivari, Hirschheim et al. 1998: 186). The most apparent difference between both theories can be found in the principles of the ISD process. Whereas DEMO strictly applies discourse analysis as means to identify the business communication of organization, LAP also applies conversation analysis. The latter is more of an empirical approach, whereas the former is more rationalist[18]. The position of the analyst in the underlying epistemological viewpoint is a plausible explanation for this difference, with DEMO adhering to positivism and LAP in general to antipositivism. Related to this observation is the apparently different level of abstraction when analyzing communication. While LAP seems concerned with identification and understanding of ordinary communication between subjects, DEMO abstracts from subjects to actors in a role concept. As such, DEMO theory provides a framework to identify high-level business communication between actors, which disregards the situatedness of communication between individual human beings.

---

[18] One dilemma within philosophy deals with the question whether knowledge is a-priori or not. Whereas rationalists claim knowledge can be obtained through *reason*, empiricists stress knowledge can only be obtained through *observation*. Of course, in practice most researchers divert from these 'ideal types' and apply a more moderate viewpoint (Kopytko 2001: 799).

| Aspect | DEMO | LAP |
| --- | --- | --- |
| *Data/information* | Constitutive at essential level. | Primarily constitutive but includes descriptive elements. |
| *Information Systems* | Social systems technically implemented. | Social systems technically implemented. |
| *Human beings* | Voluntaristic. | Dominantly voluntaristic but includes deterministic elements. |
| *Technology* | Subject to human choice. | N/A. |
| *Organizations/society* | Structural view. | Includes both structuralist and interactionist elements. |
| | | |
| *Epistemology* | Positivist regarding position of observer. | Antipositivist orientation but some positivist tendencies. |
| *Methodology* | Constructive conceptual development. | Mainly conceptual development; technical development. |
| *Role of IS Science* | Means-end oriented. | Means-end oriented. |
| *Value of IS Research* | Organizational effectiveness. | Rational and successful communication; intersubjectivity; organizational effectiveness. |

**Table 4-2: Overview of paradigmatic assumptions for both DEMO and LAP**

As the previous subsection revealed, DEMO regards the analyst as an objective observer. While this viewpoint heavily affects DEMO methodology – aiming to provide an objective blueprint of organization – its implication cannot be transferred to LAP. Although the paradigmatic differences between DEMO theory and LAP are not strikingly different besides their epistemological positions, the role of the observer is an incommensurable disparity between both theories[19]. This is an interesting observation, because by definition DEMO is also part of the LAP research community. Apparently LAP's notion of communication can be applied in various methodologies, each incorporating different paradigmatic assumptions. Therefore LAP is a pluralist aggregation of research programs and methodologies. Taking into account the observation of Iivari, Hirschheim et al. (1998: 179) about the lack of empirical research in the LAP research community, LAP research does seem to have a tendency towards rationalism, though.

## 4.3 Critical Analysis of LAP

Information Systems Development (ISD) is an applied discipline. ISD practitioners do not just engineer artifacts, but application of these artifacts affects organization itself. Formalization of work practice due to the introduction of a new Information System is a profound example. The invoked change can be regarded as an intervention. In this respect, organizational analysts are not merely passive observers, but actively involved in organizational change themselves. To possibly understand the effects of the intended intervention, full appreciation and a degree of understanding of organization as social phenomenon is required. To do so, each and every organization has to be regarded as a unique object of study. Due to the subjective interpretation of the analyst[20] the resulting impression of organization fits within the ontological position of nominalism. Nevertheless, as the term *engineering* of artifacts suggests, the design and construction of Information Systems deals with formal, constructive methods and techniques as well. As this requires a completely different approach, ISD practice has to deal with an inherent contradiction of perspectives.

The Language/Action Perspective utilizes the study of communication within organization as the basis for the design of Information Systems. The generic schema of *conversation for action* brought the speech act theory to the attention of ISD research (Ågerfalk and Eriksson 2004: 83-4). From its start, this approach has mostly been criticized by researchers from the field of ethno methodology[21] (De Michelis and Grasso 1994: 90). Most comments, however, are not targeted specifically at LAP, but more to ISD research in general. The framework of Ljungberg and Holm (1996) breaks this observation down into the following focal points:
1. The problems of theoretical abstractions
    a. The insufficiency of any theoretical abstraction;
    b. The insufficiency of particular abstractions, in this case speech act theory;
2. The problems with a rationalistic design of work (i.e. problems with rigid design versus flexibility, and global authority versus local autonomy).

---

[19] Note this is a purely theoretical observation. From a pragmatic point of view, theoretical incommensurability is not that big of an issue, as each theory is adapted to fit the problem situation at hand.
[20] As Mingers (2000: 682) points out, this observation is conditioned by previous experiences and access to the situation. This is especially relevant for agents from outside the organization.

Regarding the first point, particularly Orlikowski clarifies the dual nature of any category system, as they are both enabling and constraining (Bannon, Agre et al. 1995: 73-77). But as the previous subsection concluded, application of a category system such as speech act theory *an sich* cannot be linked to a particular paradigmatic position. The arguments seem to revolve around the already mentioned rationalist vs. empiricist debate. The second point can be traced to the dominant position of LAP regarding ethics. As summarized in the previous subsection, LAP in general adheres to a means-end orientation with respect to the role of IS Science. As such, it does not question its possible effects when applied in a specific context, but merely focuses on providing a solid and sound theory – open to any application.

The major theme in debates about the applicability of LAP for ISD seems to revolve around the classical philosophical question whether knowledge is a-priori or not. Having a tendency towards rationalism, the LAP community cannot withdraw from this debate, however. As several methodologies specifically target practitioners, LAP research is not purely theoretical. The applied paradigmatic analysis of DEMO in this paper shows it can be useful to disentangle assumptions, theories, and techniques for better understanding of a methodology. Although the applied paradigmatic framework is not without limitations itself – only to restate the dual nature of category systems, it is able to clarify several implicit aspects of LAP methodologies. As the application of DEMO by practitioners shows, there is a definite need for methods, techniques, and even methodologies that tackle a specific problem within a particular context. As LAP adheres to a means-end orientation regarding the role of IS Science, LAP can support this need of practitioners by making its intended results more clear.

### 4.4  Conclusions

To fully appreciate and understand organization as a social phenomenon, each and every organization has to be regarded as a unique object of study. The accompanied idiographic research method fits within a nominalist ontological viewpoint, which entails a subjective universe. This is completely opposite to the nature of Information Systems. Even if such systems are viewed as social systems rather than as merely technical constructions, the final system is still a formal representation of a constructed concept. This fits within the ontological position of realism. The apparent dichotomy of incompatible social and technical perspectives present in Information Systems Development is therefore rooted in paradigmatic disparity. As such, no single methodology – bound to a single paradigmatic position – can unify these incompatible perspectives. The methodologies that are part of the Language/Action Perspective are no exception.

Both the speech act theory of Austin and Searle and the communicative theory of Habermas are theories that fit within a rationalist tradition. This explains the scarcity of empirical research within the LAP community, as traditionally empiricists oppose the statement of rationalists that a-priori knowledge exists. Nevertheless, rationalism is not bound to a specific ontological or epistemological position. As such, the concepts of LAP are not particularly bound either to a single paradigmatic viewpoint. Being a pluralist aggregation of research programs and methodologies, the LAP research community supports this observation. As ISD practice is such a diverse and complex field of application, no single methodology can be expected to be suitable for each and every problem. But as an instrument to disentangle complex organization, LAP's theory about communication is promising.

---

[21] Most notable is the debate between Lucy Suchman and Terry Winograd, and the invited contributions of several other researchers (see Bannon, Agre et al. 1995).

## 5 Directions for Further Research

The conducted assessment of DEMO methodology by practitioners indicates DEMO theory has several well-developed concepts, but has a definite scope when it comes to organizational analysis. As research indicates, practitioners apply a combination of various methods, techniques, and perhaps methodologies to tackle problems. Although DEMO's positivist attitude towards the role of organizational analysts does not apply to LAP in general, some pointers to improve LAP's footprint in the ISD community of practitioners can be deduced. With respect to the role of IS Science, it should prove helpful if research programs that belong to the LAP community made their assumptions and intentions more explicit. Within a means-end orientation, a clear focus of a methodology's intended results aids practitioners in selecting an appropriate one for their task at hand. To support the combination of methodologies as applied in practice, further research into possible combinations, supported by practical interfaces, is needed. Although being a very complicated research area, the research on multi-methodologies indicates these combinations are not unattainable. Especially the voluntaristic view upon human beings and the social interpretation of Information Systems offer openings to supplementary (social) methodologies and approaches, such as those from the field of ethno methodology.

## Acknowledgements

Thanks to Rogier Krieger for critical reading and discussion of the manuscript. In addition, this paper has benefited from fruitful discussions with Jan Dietz, Sjoerd van Tuinen, Piet de Niet, and Hans Mulder. This research was partly funded by the DEMO Center of Expertise.

## Appendix 1: Elaboration of Survey Data

The complete survey contains 16 questions, of which only the relevant questions are displayed. Other questions include the amount of DEMO projects the respondent participated in, and if the respondent is willing to join the workshop. Each of the following questions is accompanied by a complete list of answers, and possibly shows some remarks about the setup. The survey was sent to 114 persons who are known to be acquainted with DEMO. The total size of this group is unknown, but is estimated to be about 1000 persons. Therefore the response of 50 (43.86%) is not significant, which implies results of the survey are only indicative.

Note: The low amount of responses from question 4 and onwards is imposed by the survey, because only respondents who participated in at least one DEMO project were asked to answer these questions.

*Q 1.* Do you mostly work for your employer (internal) or for clients (external)? (*50 responses)*.
1. Internal – 44%;
2. External – 56%.

*Q 2.* To which sectors can your clients or employer be classified? (*46 responses)*.
Remarks: The listed sectors are based upon the classification of the Dutch chamber of commerce (KvK 2004). To limit the number of answers for the respondent, the sectors are limited to six only. The automation sector was separated from the Service Industries beforehand due to the expected high share of this sector. The numbers between brackets indicate the corresponding classification numbers of the Dutch chamber of commerce.

1. Industry [15-44] – 10%;
2. Transport and Telecommunication [60-64] – 10%;
3. Service Industries [65-71, 73-74] – 27%;
4. Automation [72] – 15%;
5. Public sector (including government and education) [75-84] – 19%;
6. Various (including health and building industry)  [01-14, 45-59, 85-99] – 19%.

*Q 3*. What is the size of the organizations that commission your work? (*46 responses*).
Remarks: The sizes of the organizations are adjusted to the classification of the Dutch chamber of commerce, who distinguishes between small organizations (less than 50 employees), medium organizations (between 49 and 250 employees), and large organizations (250 or more employees).

1. 1–9 employees – 05%;
2. 10–19 employees – 02%;
3. 20–49 employees – 11%;
4. 50–99 employees – 12%;
5. 100–199 employees – 15%;
6. 200–499 employees – 13%;
7. 500–999 employees – 11%;
8. >= 1000 employees – 31%.

*Q 4.* What amount of time do you allocate for each of the following activities? (*32 responses*).

1. Outlining policy  – 08%;
2. Managing – 09%;
3. Consulting – 34%;
4. Counseling – 05%;
5. Designing – 15%;
6. Implementing – 10%;
7. Studying – 10%;
8. Teaching – 09%.

*Q 5*. For which areas of application have you used DEMO? (*32 responses*).

1. Business Process Design/Redesign – 43%;
2. Support of virtual organization – 10%;
3. Information Systems Development – 37%;
4. Other – 10%.

*Q 6*. On what level do you apply DEMO? (*31 responses*).

1. Personal level – 19%;
2. Team level – 35%;
3. Company level – 23%;
4. Commercial level – 23%.

*Q 7*. What is the average scheduled time for DEMO projects? (*28 responses*).

1. 1–3 months – 29%;
2. 4–6 months – 39%;
3. 7–12 months – 29%;
4. 1–2 years – 0%;
5. >= 2 years – 3%.

## Appendix 2: Qualitative Values of DEMO Assessment

| Dimension | Activity | Score | Variability |
|---|---|---|---|
| Social | Appreciation of | 3.4 | 65% |
| Social | Analysis of | 3.7 | 53% |
| Social | Assessment of | 4.3 | 61% |
| Social | Action to | 5.8 | 60% |
| Material | Appreciation of | 4.1 | 50% |
| Material | Analysis of | 6.5 | 58% |
| Material | Assessment of | 7.5 | 45% |
| Material | Action to | 4.4 | 52% |